# Auditory Attention Decoding from EEG using Convolutional Recurrent Neural Network

Zhen Fu[1], Bo Wang[1], Xihong Wu[1,2], Jing Chen[1,2]
[1]Department of Machine Intelligence, Speech and Hearing Research Center, and Key Laboratory of Machine Perception (Ministry of Education), Peking University, Beijing, China
[2]Peng Cheng Laboratory, Shenzhen, China
{fuzhen364, wangbo1351, xhwu, janechenjing}@pku.edu.cn

*Abstract*—The auditory attention decoding (AAD) approach was proposed to determine the identity of the attended talker in a multi-talker scenario by analyzing electroencephalography (EEG) data. Although the linear model-based method has been widely used in AAD, the linear assumption was considered oversimplified and the decoding accuracy remained lower for shorter decoding windows. Recently, nonlinear models based on deep neural networks (DNN) have been proposed to solve this problem. However, these models did not fully utilize both the spatial and temporal features of EEG, and the interpretability of DNN models was rarely investigated. In this paper, we proposed novel convolutional recurrent neural network (CRNN) based regression model and classification model, and compared them with both the linear model and the state-of-the-art DNN models. Results showed that, our proposed CRNN-based classification model outperformed others for shorter decoding windows (around 90% for 2 s and 5 s). Although worse than classification models, the decoding accuracy of the proposed CRNN-based regression model was about 5% greater than other regression models. The interpretability of DNN models was also investigated by visualizing layers' weight.

## I. INTRODUCTION

In a complex auditory scene where multiple talkers are talking simultaneously, a normal-hearing (NH) listener is usually able to selectively attend to the talker of interest while ignoring others and noise. This phenomenon is referred to as "cocktail party problem" [1], and it could be explained by the top-down attention modulation on cortical envelope tracking to speech [2]–[4]. Firstly, electroencephalography (EEG) and magnetoencephalography (MEG) studies exhibited that auditory neural activities entrain to the temporal envelope of speech [5]. Furthermore, during a selective attention task, the temporal envelope of the attended speech was more represented in listener's neural responses than that of the unattended speech, due to the top-down attentional modulation [2]–[4]. Along with these studies, the approach called auditory attention decoding (AAD) was proposed to determine the identity of the attended talker by analyzing the relation between recorded neural activities and speech envelopes. Based on the assumption of linear time-invariant system, the reconstruction filter (also called as decoder) was designed as the impulse response of the system, to map the electroencephalography (EEG) data to the attended speech envelope. The decoder was then convolved with EEG data to reconstruct the speech envelope. Consequently, the talker whose actual speech envelope has the maximum correlation coefficient with the reconstructed envelope would be determined as listener's attended target [6], [7]. It has been reported that AAD accuracies of linear models can reach a level as high as about 80–90% using 60-s decoding window, but with shortened decoding windows (e.g., 5 and 10 s), the decoding accuracy significantly dropped to about 60–70% [8], [9]. Apparently, the long decoding window is too sluggish to be used for application, e.g. in the neuro-steered hearing aids [10], [11].

Deep neural network (DNN) has advantages in nonlinear modelling and automatic feature extraction. The application of DNN in AAD has made some progresses in improving the decoding accuracy [12]–[15]. Firstly, a fully-connected network (FCN) based model was proposed to reconstruct speech envelope from EEG [14], as a replacement of the linear decoder. The decoding result (i.e., identity of the attended talker) was then determined by correlation analysis. The AAD accuracy was about 70% for a 10-s decoding window, which was greater than the linear model. Besides this regression model, DNN-based classification models were also proposed. A convolutional neural network (CNN) based model was proposed to directly determine the locus of the attended talker, which achieved about 80% accuracy with 2-s EEG data [13], [15]. In another study [12], EEG and speech envelopes were firstly transformed into a common latent space using CNN and long short term memory (LSTM) layers, respectively. The identity of the attended talker was then determined based on similarity analysis in the latent space. The AAD accuracy was about 80% for a 5-s decoding window. Generally, DNN-based models outperformed the linear model, especially for shorter decoding windows. Among these DNN-based models, CNN is the most commonly used architecture for feature extraction, due to its ability to capture the spatial features of the EEG. However, the LSTM layer which is beneficial to model sequential data and learn temporal context information, is barely adopted for feature extraction of EEG.

In order to combine the advantages of CNN and LSTM, in the current work, we proposed two novel convolutional recurrent neural network (CRNN) based models, one for the regression task and the other one for the classification task. Such an architecture was expected to learn both the spatial and temporal features of EEG. We compared the proposed models with the linear model and several state-of-the-art (SOAT) DNN-based models, and tried to investigate the interpretability of these DNN model

## II. METHODS

### A. CRNN-based Regression Model

The basic idea of our proposed CRNN-based regression model is that both the spatial and temporal context information of EEG should be taken into consideration when predicting the speech envelope. The model consists of three layers: CNN, LSTM and dense layer. The processed EEG (62 channels in the present work) was firstly fed into a causal CNN layer (5 kernels, size: 16×62) to extract spatial features. The five extracted features were then separately passed through one unique LSTM layer (4 hidden units) to extract the temporal information, resulting in five hybrid features. Finally, a shared

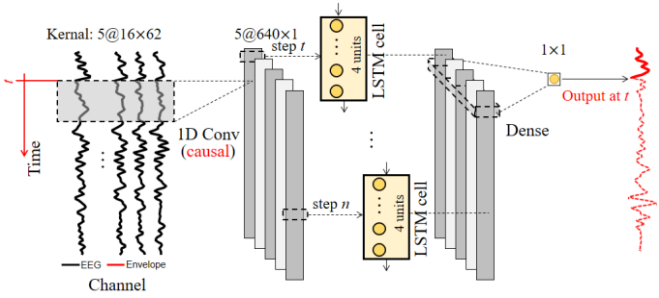

Fig. 1. Diagram of the proposed CRNN-based regression model for AAD. The figure shows the case when the decoding window length is 10 s.

dense layer (1 unit) was applied to predict the amplitude of speech envelope at each sampling time.

### B. CRNN-based Classification Model

The proposed CRNN-based classification model is an end-to-end model. Both the two speech envelopes and EEG data were provided as the input, by concatenating the two speech envelopes to the left and the right of the EEG data matrix, accordingly. The model comprises of one CNN layer (5 kernels, size: 16×64), one LSTM layer (16 hidden units), and one dense layer (2 units) that was fed to a softmax classifier. One-hot label was used to indicate the attended talker. As mentioned in [15] that a latency should be introduced between EEG and speech envelopes in consideration of the effect of attention on timings of neural responses [3], a latency of 94 ms was used when organizing the dataset.

### C. Linear Regression Model

The linear regression model used here is the same as in [6]–[9]. The decoder was calculated by the reverse correlation method with regularization. Following the parameter settings in , time lags ranged from 0 to 500 ms post-stimulus and the regularization parameter was determined according to a leave-one-out cross-validation approach.

### D. FCN-based Regression Model.

The FCN model used in the present work is the same as that proposed in [14] for speech envelope reconstruction, which served as the SOAT regression model. Specifically, in the training scheme, the EEG data within a certain temporal context (27 points, corresponding to about 422 ms) were fed to two dense layers (2 and 1 unit, respectively) to predict the amplitude of speech envelope in a sample-wise way. In other words, continuous speech envelopes were predicted using the same FCN weights.

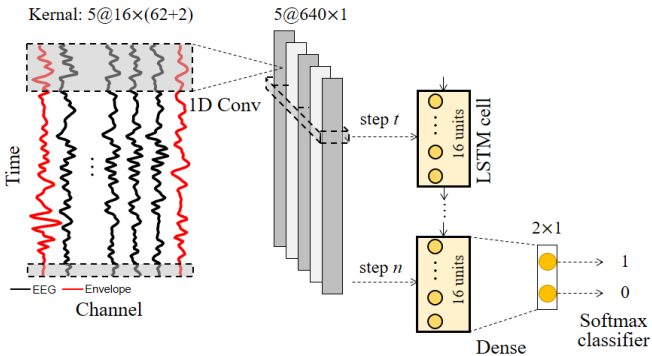

Fig. 2. Diagram of the proposed CRNN-based classification model for AAD. The figure shows the case when the decoding window length is 10 s.

### E. CNN-based Classification Model.

The CNN-based classification model used in the present work is the same as that proposed in [15]. The input of the model was the same as our proposed CRNN-based classification model. The model comprised of one CNN layer (5 kernels, size: 9×64), one average-pooling layer and two dense layers (5 and 2 units, respectively) that were fed to a softmax classifier.

### F. CNN+LSTM-based Classification Model.

The model used here was an variant of the model proposed in [12] so that it was suitable for a two-talker scenario. And this model served as the SOAT classification model. The two speech envelopes and EEG data were passed through two separate networks, respectively. The two envelopes were fed to a shared network that consisted of a 1-D CNN layer (16 kernels, kernel size: 9×1) and a LSTM layer (16 hidden units). The EEG data were fed to another network that comprised of a 2-D CNN layer (8 kernels, kernel size: 9×7) and two-dense layer (20 and 16 units, respectively). The cosine similarity of the EEG feature and each envelope feature were calculated respectively. Finally, a softmax classifier was applied to determine which speech was attended.

## III. EXPERIMENTS

### A. Participants and Materials

EEG data were recorded from twelve Mandarin-native NH listeners. All subjects were given informed consent approved by the Peking University Institutional Review Board.

The stimuli used in this experiment were the audiovisual speech corpus developed in [8]. There were 18 pairs of 60-s length audiovisual speech and each pair contained materials of a female and a male talker. The experiment was performed in an acoustically and electrically shielded booth. The audiovisual (AV) stimuli were presented with two pairs of screen-loudspeaker on the front left and front right of the subject symmetrically. The screen-loudspeaker were 0.8 meters away from subjects, with three spatial separations from the subject, 30°, 60° and 90°. During the presentation, the female AV material was presented by one pair of the screen-loudspeaker and the male AV material was presented by another pair. Subjects were instructed to pay both auditory and visual attention to the target talker, which was cued with a red frame on either of the two screens. The target talker would be switched between the two talkers for every 15 s or 30 s. Subjects could saccade and rotate head freely during the stimuli presentation. Therefore, there were 6 conditions (2 switching intervals × 3 spatial separations). And for each subject, there were 3 pairs of AV speech out of 18 assigned to each condition, and no AV speech was repeated across conditions.

Continuous EEG data were acquired by a NeuroScan system, using 62 scalp, 4 electrooculograms (EOG), and 4 electromyography (EMG) electrodes, with the reference on the nose-tip. The recording of EOG and EMG was for the purpose of artifacts rejection. EEG data were online bandpass filtered (0.15–100 Hz), digitized (500 Hz sampling rate), and stored for offline analysis. For each subject, 18 pairs of audiovisual speech were presented twice with different switching targets, thus there were 36 trials per subject and 432-min data (=36 trials × 12 subjects) in total.

## B. Speech and EEG Signal Preprocessing

To calculate the speech temporal envelope, the 60-s clean speech signal was first passed through an auditory bandpass filter-bank. Hilbert transform was then performed on the output of each band. The analytical signals were then power-law compressed and lowpass filtered. Lastly, the filtered signal of all bands were averaged to obtain the temporal envelope. As suggested in [14], [15], the cutoff frequency of the lowpass filtering was 8 Hz for use with the linear model, and was 32 Hz for use with DNN models.

EEG data were processed using the EEGLAB toolbox [16]. After recalculating noisy channels by interpolation, EEG data were re-referenced to a common average reference, downsampled to 64 Hz, baseline corrected and lowpass filtered (the cutoff frequency was matched with the speech envelope calculation). Then, artifacts (e.g., saccade, blink and head rotation) were removed by using independent component analysis (ICA). After all, both speech temporal envelope and EEG signals were normalized to have 0 mean and 1 variance.

## C. Experiment Setup

For each subject, data of the first 28 trials and the last 8 trials (corresponding to the first 14 and the last 4 pairs of AV stimuli) were allocated to the training and testing set, respectively. Thus, there was no overlap between the training set (77.8%, 336 mins) and testing set (22.2%, 96 mins). Afterwards, both sets were further split into segments with duration of 2, 5 and 10 s with 50% overlapping. In summary, there were three different decoding windows (2, 5 and 10 s) adopted for both training and testing in the present study.

For DNN-based regression models, the envelope of the actual attended speech served as the supervision, and the correlation coefficient (Pearson's $r$) between the predicted envelop and the actual attended envelope was calculated as a measure of loss function [14]. For DNN-based classification models, the identity of the actual attended talker served as supervision, and the binary cross-entropy loss function was used.

During training, the Adam optimizer was adopted. To prevent overfitting, dropout rate of 25% was used, and batch size was set as 128. Training was stopped when no loss reduction was found for 10 consecutive training epochs. During testing, for regression models, the classification result was determined by selecting the talker whose actual speech envelope had the greater correlation coefficient with the reconstructed envelope. And classification models directly output the identity of the attended talker. The AAD accuracy was further evaluated by the percentage of correctly-classified trials among all testing trials.

The linear model was implemented by the mTRF toolbox [17] and the DNN-based models were implemented based on Keras 2.3.1 platform [18].

## IV. RESULT AND DISCUSSION

### A. Decoding Performance

The model complexity and decoding accuracy for each model as a function of decoding window are shown in Table 1. As expected, as the simplest model, the linear model performed the worst among all models. For DNN-based classification models, our proposed CRNN-based model outperformed others for shorter decoding windows (i.e., 2 and 5 s), indicating the effectiveness of combining both spatial and temporal features. CNN-based classification model was the best for 10-s decoding window. For the three regression models, our proposed CRNN-based model outperformed the FCN-based model and was the best for all decoding windows. For the four existing models (i.e., linear, FCN, CNN and LSTM+CNN), AAD performances calculated in the present study were quite similar to the results reported in literatures [12]–[15].

TABLE I. MODEL COMPLEXITY AND DECODING ACCURACY OF EACH MODEL. BOLDFACE INDICATED BEST RESULT.

|  | Model | #Para. | Window (s) | | |
| --- | --- | --- | --- | --- | --- |
|  |  |  | 2 | 5 | 10 |
| Classification | CNN [15] | 2.7 k | 83.8 | 84.2 | **81.2** |
|  | CNN+LSTM [12] | 7.5 k | 82.3 | 80.6 | 66.4 |
|  | CRNN | 6.2 k | **87.4** | **90.9** | 72.9 |
| Regression | Linear | 2 k | 56.9 | 60.4 | 60.8 |
|  | FCN[14] | 3.1 k | 56.2 | 60.9 | 62.3 |
|  | CRNN | 5.2 k | **61.3** | **66.6** | **67.6** |

The results also showed that classification models outperformed regression models for most conditions. This is likely because classification models directly optimize the AAD accuracy, but regression models only optimize the accuracy of reconstruction. However, it should be mentioned that the CNN-based and CRNN-based classification models works well for binaural scenarios, but not for monaural scenarios, since the two models mainly decodes the locus of spatial attention [19]. On the contrary, regression models determine auditory attention by comparing the similarity between the reconstructed envelope and actual envelopes, so it works for both binaural scenarios and monaural scenarios. Although the CNN+LSTM based classification model outputs auditory attention directly, it actually does regression in the latent space, which makes it working on both scenarios as well.

To explore the robustness of different models to attention switching, continuous decoding results of testing trials are shown in Fig. 3, with a 5-s decoding window and 15-s switching interval as an example. On average, all models could correctly respond to the attention switching with a latency of about 5 s. DNN-based models generally had shorter latency and less errors than the linear model. Besides, comparing with regression models, classification models were also more sensitive to the attention switching. These results preliminarily indicated the feasibility of applying the AAD methods in realistic scenarios in which attention switching happens frequently. Nevertheless, reduction of the algorithm latency is still challenging and worthy of further research.

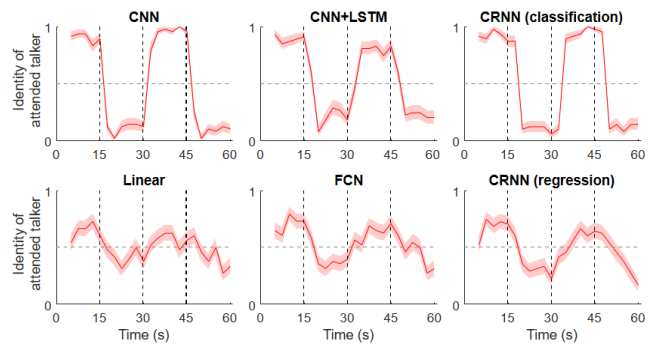

Fig. 3. Continuous decoding results averaged across testing trials. Vertical dashed lines indicates the timings of attention switching, and the shaded area indicates standard error. The figure shows the case that the decoding window is 5 s and the switching interval is 15 s.

## B. Interpretability of DNN Models

For the linear model, the decoder is considered as a set of spatial-temporal filters, and the value of each filter could be considered as the weighting parameter at certain time lags [7]–[9]. For example, both temporal and occipital region were found contributing to AAD prominently for AV condition, and there was a difference between the timings of these two regions involving in AAD [8]. However, the interpretability of DNN models is considered quite challenging. In the present study, we attempted to interpret these AAD models by visualizing the weights of DNN layers. For the CNN and CRNN-based models, only the 1-D CNN layer was visualized for the convenience of illustration. For the FCN-based regression model, the dense connections were treated as kernels. The linear decoder was also visualized for comparison.

As suggested in [8], [20], the averaged and RMS value across time were used to describe the general characteristics of each kernel. The visualizations are shown in Fig. 4. For the CNN-based classification model, the first 3 kernels were consistent across time (indicated by the RMS value), and had opposite polarities between the left and right frontal lobe (indicated by the averaged value). Therefore, these kernels were considered relating to the spatial attention to the target talker. The last 2 kernels had similar averaged patterns with others, but were inconsistent across time over temporal and occipital region. Hence, besides the spatial attention, these kernels were also related to temporal processing (i.e., processing AV stimuli). For the proposed CRNN-based classification model, the patterns of kernels were more complex than the former model. All kernels exhibited the opposite polarities between the left and right frontal lobe, and were inconsistent across time as well, which indicated that the kernels learned to extract features of both spatial and temporal dimensions. Kernels of regression models (i.e., CRNN-based, FCN-based and the linear model), unlike the classification models, all showed prominent contributions from temporal and occipital regions. These patterns were consistent with the linear decoder reported in [8]. This is rational because regression models were built to learn the mapping from EEG to temporal envelope of speech, and both temporal and occipital regions were demonstrated processing the speech envelope for the congruent AV condition [8]. Relative to the linear model, the more complex kernels of the two DNN-based regression models indicated that DNN models learned more sophisticated mapping systems. This could be the reason that DNN-based regression models outperformed the linear model.

The ability to further extract the temporal features by the LSTM layer amid the two CRNN-based models, is likely account for their higher decoding accuracy than other models of the same type. However, the visualization of weights of LSTM layer is not intuitive. Therefore, a thorough discussion about the interpretability of complicated DNN models is still challenging.

## V. CONCLUSION

In this paper, we proposed two novel CRNN-based models for the AAD task, from the perspective of regression and classification. To verify the validity of our proposed models, we compared the AAD performances with the linear model and SOAT DNN-based models, and three types of decoding windows. Furthermore, the interpretability of these DNN-based models was investigated by visualizing the weights of DNN layers. The main findings were, (1) our proposed CRNN-based models both outperformed other models of the same type, especially with shorter decoding windows; and (2) our proposed CRNN-based models had a certain degree of interpretability, since they were demonstrated extracting both spatial and temporal feathers of EEG. The improvements of AAD performance by our novel DNN-based models favored the possible AAD application, such as neuro-steered hearing aids.

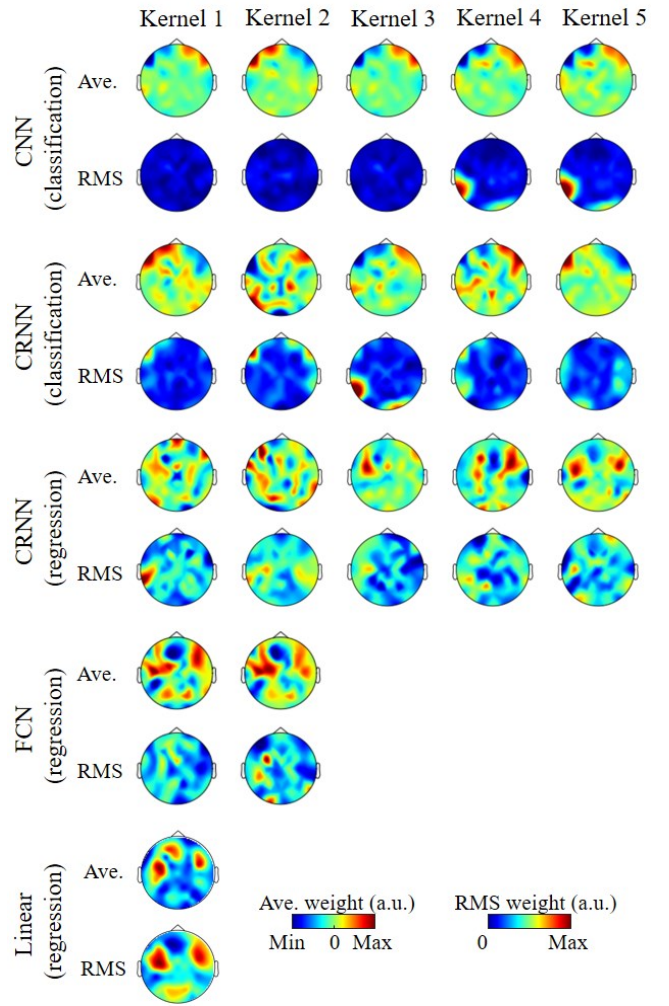

Fig. 4. Topographies of the averaged and RMS value of layer weights for different models.

## ACKNOWLEDGMENT

This work was supported by the National Natural Science Foundation of China (Grant Nos. 61771023, 12074012, and U1713217), a research funding from SONOVA, and the High-performance Computing Platform of Peking University.

## REFERENCES

[1] E. C. Cherry, "Some experiments on the recognition of speech, with one and with two ears," J. Acoust. Soc. Am., vol. 25, no. 5, pp. 975–979, Sep. 1953.

[2] H. Luo and D. Poeppel, "Phase patterns of neuronal responses reliably discriminate speech in human auditory cortex," Neuron, vol. 54, no. 6, pp. 1001–1010, Jun. 2007.

[3] N. Ding and J. Z. Simon, "Emergence of neural encoding of auditory objects while listening to competing speakers," Proc. Natl. Acad. Sci. U.S.A., vol. 109, no. 29, pp. 11854–11859, Jul. 2012.

[4] N. Mesgarani and E. F. Chang, "Selective cortical representation of attended speaker in multi-talker speech perception," Nature, vol. 485, no. 7397, pp. 233–236, May 2012.

[5] E. C. Lalor and J. J. Foxe, "Neural responses to uninterrupted natural speech can be extracted with precise temporal resolution," Eur. J. Neurosci., vol. 31, no. 1, pp. 189–193, Jan. 2010.

[6] B. Mirkovic, S. Debener, M. Jaeger, and M. De Vos, "Decoding the attended speech stream with multi-channel EEG: implications for online, daily-life applications," J. Neural Eng., vol. 12, no. 4, p. 046007, Aug. 2015.

[7] J. A. O'Sullivan et al., "Attentional selection in a cocktail party environment can be decoded from single-trial EEG," Cereb. Cortex, vol. 25, no. 7, pp. 1697–1706, Jul. 2015.

[8] Z. Fu, X. Wu, and J. Chen, "Congruent audiovisual speech enhances auditory attention decoding with EEG," J. Neural Eng., vol. 16, no. 6, p. 066033, Nov. 2019.

[9] S. A. Fuglsang, T. Dau, and J. Hjortkjær, "Noise-robust cortical tracking of attended speech in real-world acoustic scenes," Neuroimage, vol. 156, pp. 435–444, Aug. 2017.

[10] L. Fiedler, J. Obleser, T. Lunner, and C. Graversen, "Ear-EEG allows extraction of neural responses in challenging listening scenarios — A future technology for hearing aids?," in 2016 38th Annual International Conference of the IEEE Engineering in Medicine and Biology Society (EMBC), Orlando, FL, USA, Aug. 2016, pp. 5697–5700.

[11] W. Pu, P. Zan, J. Xiao, T. Zhang, and Z.-Q. Luo, "Evaluation of joint auditory attention decoding and adaptive binaural beamforming approach for hearing devices with attention switching," in ICASSP 2020 - 2020 IEEE International Conference on Acoustics, Speech and Signal Processing (ICASSP), Barcelona, Spain, May 2020, pp. 8728–8732.

[12] M. J. Monesi, B. Accou, J. Montoya-Martinez, T. Francart, and H. V. Hamme, "An LSTM based architecture to relate speech stimulus to EEG" in ICASSP 2020 - 2020 IEEE International Conference on Acoustics, Speech and Signal Processing (ICASSP), Barcelona, Spain, May 2020, pp. 941–945.

[13] G. Ciccarelli et al., "Comparison of two-talker attention decoding from EEG with nonlinear neural networks and linear methods," Sci Rep, vol. 9, no. 1, p. 11538, Dec. 2019.

[14] T. de Taillez, B. Kollmeier, and B. T. Meyer, "Machine learning for decoding listeners' attention from electroencephalography evoked by continuous speech," Eur. J. Neurosci., vol. 51, no. 5, pp. 1234–1241, Mar. 2020.

[15] L. Deckers, N. Das, A. H. Ansari, A. Bertrand, and T. Francart, "EEG-based detection of the locus of auditory attention with convolutional neural networks," Neuroscience, preprint, Nov. 2018.

[16] A. Delorme and S. Makeig, "EEGLAB: an open source toolbox for analysis of single-trial EEG dynamics including independent component analysis," J. Neurosci. Methods, vol. 134, no. 1, pp. 9–21, Mar. 2004.

[17] M. J. Crosse, G. M. Di Liberto, A. Bednar, and E. C. Lalor, "The multivariate temporal response function (mTRF) toolbox: a MATLAB toolbox for relating neural signals to continuous stimuli," Front. Hum. Neurosci., vol. 10, Nov. 2016.

[18] F. Chollet and others, "Keras," 2015. https://keras.io.

[19] S. Vandecappelle, L. Deckers, N. Das, A. H. Ansari, A. Bertrand, and T. Francart, "EEG-based detection of the locus of auditory attention with convolutional neural networks," Neuroscience, preprint, Feb. 2020.

[20] Z. Fu and J. Chen, "Congruent audiovisual speech enhances cortical envelope tracking during auditory selective attention," in Interspeech 2020, Oct. 2020, pp. 116–120.